# Galaxy Mergers and Gravitational Lens Statistics

Hans-Walter Rix[1] and Dan Maoz[2]

Institute for Advanced Study, Princeton, NJ 08540

Edwin L. Turner

Princeton University Observatory, Princeton, NJ 08540

and

Masataka Fukugita[3]

Institute for Advanced Study, Princeton, NJ 08540

astro-ph/9405014

[1]Hubble Fellow

[2]Present address: School of Physics and Astronomy and Wise Observatory, Tel-Aviv University, Tel-Aviv 69978, Israel

[3]Also at Yukawa Institute, Kyoto University, Kyoto 606, Japan

# ABSTRACT

We investigate the impact of hierarchical galaxy merging on the statistics of gravitational lensing of distant sources. Since no definite theoretical predictions for the merging history of luminous galaxies exist, we adopt a parametrized prescription, which allows us to adjust the expected number of pieces comprising a typical present galaxy at $z \sim 0.65$. The existence of global parameter relations for elliptical galaxies and constraints on the evolution of the phase space density in dissipationless mergers, allow us to limit the possible evolution of galaxy lens properties under merging.

We draw two lessons from implementing this lens evolution into statistical lens calculations: (1) The total optical depth to multiple imaging (e.g. of quasars) is quite insensitive to merging. (2) Merging leads to a smaller mean separation of observed multiple images. Because merging does not reduce drastically the expected lensing frequency it cannot make $\lambda$-dominated cosmologies compatible with the existing lensing observations. A comparison with the data from the HST Snapshot Survey shows that models with little or no evolution of the lens population are statistically favored over strong merging scenarios. The specific merging scenario proposed by Toomre (1977) can be rejected (95% level) by such a comparison. Some versions of the scenario proposed by Broadhurst, Ellis & Glazebrook (1992), are statistically acceptable.

## 1. Introduction

The merging of galaxies and their halos is an integral part of every hierarchical structure formation scenario. Such mergers are not only expected, but also observed directly in the nearby universe. Galaxy mergers and gravitational lens statistics are connected because elliptical galaxies – the merger products – constitute the most important galaxy lens population at low redshifts.

Although the occurrence of galaxy mergers is no longer a matter of dispute, there is no agreement on the current and past rate of galaxy mergers and on their importance in the formation of elliptical galaxies. It has been suggested (see Schweizer, 1990, for a review) that most luminous elliptical galaxies formed at $z \lesssim 1$ via dynamical merger of other galaxies, often taken to be gas-rich disk galaxies. This hypothesis draws support from N-body simulations (see Barnes & Hernquist, 1992, and references therein) and from observations of low redshift galaxy interactions (e.g. Schweizer 1990). Although estimates of the current galaxy merger rate are low, a moderately higher rate in the past could have given rise to most or all present day ellipticals (Toomre 1977, Schweizer, 1990). The simplest versions of this hypothesis, however, face serious difficulty in accounting for many of the observed dynamical and statistical properties of ellipticals (e.g. Ostriker, 1980, Carlberg 1986).

Recently, the possibility that mergers have greatly modified the galaxy population during recent epochs ($z \lesssim 0.5 - 1$) has received attention from a new direction: one way to understand the results from counts and redshift surveys of very faint galaxy populations is to assume that the universe at $z \sim 0.5$ contained many more medium to low luminosity galaxies than it does now (Broadhurst, Ellis and Glazebrook, 1992, hereafter BEG, and references therein); the excess population of small galaxies has disappeared by merger into larger, presumably elliptical, systems (BEG, but see Dalcanton, 1994).

Adopting the viewpoint that the amount of galaxy merging since $z \sim 1$ is an open question, we explore in this paper the impact of galaxy mergers, forming elliptical galaxies, on gravitational lens statistics. Merging may be important to lensing for two reasons: first, the gravitational lensing rates, particularly for events with image separations of more than an arc second, presumably are dominated by early-type galaxies (e.g. Turner, Ostriker & Gott 1984, hereafter TOG; Fukugita & Turner 1991, hereafter FT; Kochanek 1991). Second, the same calculations indicate that quasar lensing rates are dominated by galaxies at redshifts between $\sim 0.2$ and $\sim 1$, corresponding to large fractional look back times.

Most lensing calculations (see Narayan and White, 1988, Mao, 1991, for exceptions) have only considered non-evolving galaxy populations and have found satisfactory agreement with observations. It is conceivable that at least extreme versions of the merger hypothesis might spoil this agreement and could thus be excluded. In particular, a detailed analysis by Maoz and Rix (1993, hereafter MR) of the best available statistical sample of lenses, the HST Snapshot Survey (Maoz *et al.*, 1993 and references therein), shows that a conventional model of elliptical galaxies with massive dark halos reproduces the observations well within a standard cosmological model (see also Kochanek, 1993). Moreover, MR demonstrated that the data are inconsistent with large deviations from these conservative assumptions (e.g., absence of dark halos or large values of the cosmological constant). Thus, the lensing data can provide significant cosmological and cosmogonic constraints.

In Section II we describe a simple parameterization of the rate and dynamical character of galaxy mergers. In Section III we first illustrate some basic effects using a singular isothermal sphere (SIS) lens model; in Section IV we model the lenses as early-type galaxies and statistically compare merger model predictions to the HST Snapshot Survey data following the techniques of MR. In Section V we focus on the question of whether allowing for lens evolution through merging can weaken the limits on the cosmological constant $\lambda$, derived for no-evolution models. Finally, in Section VI we state the conclusions from these calculations.

While this paper was in a late draft stage, we received a preprint by Mao and Kochanek (1993) which reports a somewhat similar investigation and arriving at qualitatively similar results. Their study considers a much wider variety of types of galaxy evolution than mergers but treats merger models themselves in less detail and variety. Section VI contains some further comparison of our conclusions to theirs.

## 2.   Parameterization of Galaxy Merger Scenarios

At each cosmic epoch, $z$, galaxies and their properties enter into statistical lensing calculations through their contribution to the optical depth to image splitting at a given separation. This contribution depends on (1) their number density and mass function and (2) on their surface mass density profile, both of which are affected by merging.

The evolution of the galaxy merging rate and the details of individual galaxy mergers are quite complex (e.g. Lacey and Cole, 1993, Kauffmann and White, 1993), and no reliable prediction for the merger rates of luminous galaxies exist. Therefore we propose a simple parameterization, separating the two just mentioned aspects of galaxy evolution through merging.

## 2.1.   How Much Merging?

For the temporal evolution of the galaxy number density we adopt the parametrized, *ad hoc* prescription suggested by BEG to match the counts and redshifts of faint galaxies. These authors suggest that the galaxy mass function $\Phi^*$ at look-back time $\delta t$ may be

written as

$$\Phi^\star(\delta t, M) = f(\delta t)^2 \Phi_0^\star(f(\delta t) \times M), \tag{1}$$

where the subscript 0 indicates the $z = 0$ value. The time dependence enters through

$$f(\delta t) = \exp\left\{Q\delta t/\beta t_0\right\} = \exp\{-Q[(1+z)^{-\beta} - 1]/\beta\}, \tag{2}$$

where $\beta$ is the ratio of the Hubble time $H_0^{-1}$ to the age of the Universe $t_0$ (e.g. $\beta = 1.5$ for $\Omega_0 = 1$; note that the right-hand side equality in equation [2] is invalid for models with a cosmological constant, to be discussed in §5). This prescription (1) can be visualized as a "family tree", going back in time, in which each galaxy splits into two pieces whenever the value of $f$ doubles. The merging rate $Q$ roughly corresponds to the average number of pieces at $z \sim 0.65$ (near the peak of the expected *lens* redshift distribution for typical QSO samples) that merged to produce one galaxy at $z = 0$. BEG find that in an $\Omega_0 = 1$ universe the observations of faint blue galaxies can be matched by $Q \sim 4$.

## 2.2. How Do Galaxy Properties Change under Merging?

The galaxy property most important to lensing statistics is the enclosed mass profile, $M(< R)/R$, which for axisymmetric lenses determines the gravitational deflection angle $\alpha$,

$$\alpha(R) = \frac{4GM_{<R}}{c^2 R}. \tag{3}$$

This angle in turn determines the total cross section for multiple imaging at given image separations. It is therefore important to model how merger events redistribute mass within the coalescing galaxies, which is constrained both theoretically and observationally, as we will now show.

As in MR, we consider the well studied luminous part of the galaxies and their dark halos separately. Early-type galaxies now occupy a "fundamental plane" in the three-dimensional space of parameters ($\sigma_{central}, L, R_{eff}$) describing their global properties (Kormendy and Djorgovski 1989, see also MR). Merging removes two galaxies located in this parameter space and places the virialized merger product at a different place in this space. The existence of a fundamental plane at present is most natural, if merging moves galaxies *within* the fundamental plane. Since a galaxy's luminosity narrowly confines the other two parameters, the fundamental plane provides a unique way of assigning new parameters to the stellar body of the merger product. In this scenario the merger product is less concentrated than its progenitors, since the effective galaxy surface brightness is observed to decrease with increasing luminosity (see MR).

This purely empirical argument is supported by numerical simulations: Okumura, Ebisuzaki & Makino (1991) and Hernquist, Spergel & Heyl (1993), show that, in the inner parts of a galaxy, the decrease in mean phase space density during a merger is close to that required for moving within the fundamental plane. Since the majority of the stars in elliptical galaxies is likely to have formed at a redshift $\gtrsim 1$ (Bower, Lucey &Ellis 1992; Aragón-Salamanca *et al.* 1993), all mergers since then should be simulated well in (dissipationless) N-body experiments.

To describe the merging of dark halos, we resort to a simple parameterization, assuming at this point only that the dark matter is dissipationless. We introduce this parameterization using isothermal spheres, because we will assume in the final models that the halos are isothermal (but with a core). For these models merging can be characterized by an index $\nu$ defined through

$$\sigma(M, z) = f\left[\delta t(z)\right]^{-\nu} \sigma_0\left[f\left[\delta t(z)\right] M\right], \tag{4}$$

where $\sigma_0(M)$ is the velocity dispersion associated with a halo of mass $M$ at $z = 0$ and $\sigma(M, z)$ is the velocity dispersion associated with mass $M$ at redshift $z$. Thus the merger product of $f(z)$ earlier galaxies is taken to have a velocity dispersion that is a factor of $f(z)^{\nu}$ greater than that of its progenitors. For SIS, the merged halo produces a bend angle that is $f(z)^{2\nu}$ times larger and a cross section to lensing that is typically $f(z)^{4\nu}$ times larger than one of its average progenitors. For $\nu = 1/4$ both pre-merger and post-merger are on the fundamental plane.

Different values of $\nu$ can be related to the evolution of the phase space density in the course of a merger and thus, at least loosely, to assumptions about the underlying merger physics. For an isothermal sphere of dispersion $\sigma$, the phase space density, $\psi(M)$, of a thin shell enclosing a mass of $M$ is

$$\psi(M) \propto \frac{\sigma^3}{M^2}. \tag{5}$$

Using the cumulative mass as the radial coordinate allows us to compare conveniently the pre-merger phase space densities, $\psi(M)$, with the post-merger value $\psi(2M)$ (for mergers of equal size objects).

The change in $\sigma$ in dissipationless mergers (Eq. 4) can be bracketed by $\nu = 0$ and $\nu = 2/3$. The lower limit arises from the virial theorem and energy conservation: for objects merging without dissipation from near-parabolic orbits the total energy per unit mass is conserved. In virialized systems this energy is $-\langle \sigma_{before}^2 \rangle = -\langle \sigma_{after} \rangle$. All the energy gained in the coalescence is spent on expanding the merger product to the same specific binding energy. With Equation 5 we find that $\psi$ increases by a factor of four for each doubling of the mass. Note, however, that this argument only pertains to the system as a whole and need not apply to the inner parts of the merger product which determine gravitational lensing properties. The other extreme follows from the assumption that the merging process mixes no "empty phase space" into the coarse grain distribution function. Since $\psi$ cannot increase, $\psi = const$ implies $\sigma \propto M^{2/3}$ (see Equation 5), corresponding to $\nu = 2/3$.

There are good reasons to believe that $\nu$ is close to $1/4$. For merging of isothermal spheres of equal mass scale, this corresponds to a dilution of the coarse grained phase space density of $\sim 2.5$ during a merger. N-body simulations (Okumura *et al.* 1991; Hernquist, Spergel and Heyl, 1993) have shown that merging from near parabolic orbits causes the mean phase space density of the most tightly bound half of the material to decrease by a factor of $2 - 4$, independent of the details of the merger orbit or the radial distribution of the material.

In order to make ellipticals out of spirals, the merging process must increase the resulting halo velocity dispersion dramatically. Since energy conservation yields $\nu = 0$, this implies enormous dissipation or redistribution of specific energy during the merger, a long recognized difficulty for the spirals-to-ellipticals hypothesis (Ostriker 1980; see however, Hernquist 1993). Here we model the spiral-to-elliptical scenario (Toomre 1977) by assuming that the bulges of the spiral galaxies move on the fundamental plane to become the luminous parts of the resulting elliptical galaxy, and the halos merge such that their velocity dispersions obey $\nu = 2/3$. In order to test critically the spiral-to-elliptical merging scenario, which greatly reduces the predicted number of lensed quasars, we will assume that all the luminous mass in the early-type galaxies originated in the bulges of the spiral galaxies. This conservative assumption corresponds to minimizing the attenuation of the lensing frequency due to merging.

With the parameterization developed above, we are now prepared to investigate the implications of various merger scenarios for QSO lensing statistics. We will examine values of $Q$ in the range of 2 to 4 (with $Q = 0$ giving the reference "no merging" case) and $\nu$ values between 0 and $2/3$, with a particular focus on $\nu = 1/4$.

### 3. Lensing by Merging Singular Isothermal Spheres: An Illustration

The modeling of galaxy mass distributions as singular isothermal spheres (SIS) has been a common approximation in the literature of lensing statistics (Turner 1980; TOG, Kochanek 1991, but see MR for its shortcomings). Here we will use the SIS approximation to illustrate the main, competing effects that merging has on lensing statistics:

(1) There were more galaxies, hence more lenses, in the past.

(2) Galaxies were typically less massive in the past and thus less powerful deflectors.

(3) Merging decreases the central surface mass density, while dissipation and star formation at the galaxy's center can offset this effect.

While the first effect is modeled by the merging prescription of Eqs. (1)–(2), items 2 and 3 cannot be addressed separately in the context of the SIS model, but are collectively described by the index $\nu$ (see Section 2.2).

For SIS models, expressions for the gravitational lensing optical depth, mean image separation, and mean lens redshift can be derived with an explicit dependence on $f(Q)$ and $\nu$. Their evaluation, however, still requires a one-dimensional integral. Note also, that these expressions neglect to account for the detectability of multiple imaging in actual surveys. We adopt the notation and formulation of TOG and adopt an $\Omega_0 = 1$ filled beam cosmology; the formulae themselves can be derived from those given in TOG. Denoting the lens redshift by $x \equiv 1 + z_{lens}$ and the source redshift by $y \equiv 1 + z_{QSO}$ , we have for the optical depth, $\tau$

$$\tau = 4F_0 \int_1^y \frac{\left(x^{1/2} - 1\right)^2}{x^{7/2}} \cdot \left(\frac{y^{1/2} - x^{1/2}}{y^{1/2} - 1}\right)^2 \exp\left[-\frac{Q}{\beta}\left(x^{-\beta} - 1\right)\right]^{1-4\nu} dx, \qquad (6)$$

for the mean image separation

$$\overline{\Delta\theta} = \frac{8\alpha_0^{\star}F_0}{\tau} \int_1^y \frac{\left(x^{1/2}-1\right)^2}{x^4} \left(\frac{y^{1/2}-x^{1/2}}{y^{1/2}-1}\right)^3 \exp\left[-\frac{Q}{\beta}\left(x^{-\beta}-1\right)\right]^{1-6\nu} dx, \qquad (7)$$

and for the mean lens redshift

$$\overline{z_{\mathrm{lens}}} = \frac{4F_0}{\tau} \int_1^y \frac{\left(x^{1/2}-1\right)^2(x-1)}{x^{3/2}} \left(\frac{y^{1/2}-x^{1/2}}{y^{1/2}-1}\right)^2 \exp\left[-\frac{Q}{\beta}\left(x^{-\beta}-1\right)\right]^{1-4\nu} dx, \qquad (8)$$

where $\alpha_0^{\star} = 4\pi\sigma^{\star 2}$ is the bending angle for an $L^{\star}$ galaxy and $F_0 = 16\pi^3 n_0 H_0^{-3}\sigma^{\star 4}$ ($n_0$ is the number density of galaxies) is effectiveness of galaxies producing multiple images (TOG).

Equations (6) and (8) show that for $\nu = 1/4$ the mean lens redshift, the differential (and total) optical depth at each redshift remain unchanged under merging. For this value of $\nu$ the competing effects of a smaller mean cross section per lens and a larger number density of lenses cancel. For smaller values of $\nu$, the optical depth increases, and *vice versa*. In contrast, the mean image separation remains invariant under merging for $\nu = 1/6$ (Eq. 7). This index differs from Eq. 6 and 8, because the integral in Eq. 9 contains an additional power of $\overline{\Delta\theta}(z)$ in the integrand.

The above formulae are useful for developing some intuition and obtaining quick estimates of the effects that lensing may have on lens statistics. For a more detailed and quantitative comparison with observations, we follow MR and consider the separation distribution, $n(\theta)$, that a model predicts for the QSOs in the HST Snapshot Survey (see Maoz *et al.* 1993, and references therein). This distribution represents the expected number of multiply imaged QSOs (per arcsecond interval of separation) at separation $\theta$ when looking along the lines-of-sight to the survey QSOs. We assume a present day luminosity distribution as detailed in MR, assume that the velocity dispersion parameter $\sigma_*$ for an $L_*$ galaxy is 225km s$^{-1}$ (without a $\sqrt{3/2}$ factor) (FT)[4], and assume that $\sigma$ varies as $L_*^{\frac{1}{4}}$ at present. For the no-evolution case, $n(\theta)$ is calculated by integrating over all lines of sight to the QSOs in the HST Snapshot Survey, over all possible lens redshifts, over all galaxies at a given lens redshift, and over all impact parameters (see e.g. TOG, MR, Kochanek 1993). We also take into account the ability of the HST Snapshot Survey to detect multiple images of a given separation. The inclusion of merging means that both $\Phi(L)$ and $\sigma(L)$ become functions of the lens redshift through Eqs. (1), (2) and (4).

---

[4]This value is different from the value for the *halo* velocity dispersion, 300km s$^{-1}$, of an $L_*$

Figure 1 compares $n(\theta)$ for the no-merging case (thick solid line) with a variety of merging scenarios and two different merging rates. The three other lines correspond to merging scenarios of different character, described in Section 2.2 and labeled by the parameter $\nu$. This plot complements Eqs. (6)-(8) in illustrating some of the generic features that galaxy merging may have on gravitational lens statistics:

(1) Galaxy merging skews $n(\theta)$ towards smaller separations, because galaxies (and hence lenses) were less massive in the past than they are now. If the total cross section decreases with cosmic time, $\nu > 1/4$, then a larger fraction of lensing events arises from more distant lenses, further decreasing the expected image separations.

(2) Merging can increase or decrease the expected number of lensed sources, depending on how the mean phase space density of the galaxies changes under merging. For $\nu = 1/4$, the total lensing cross section, and hence the total lensing rate, remains unchanged.

(3) If merging takes place so that it does not move galaxies in the fundamental plane (i.e. $\nu \neq 1/4$), even a modest amount of merging may change the predictions for lensing statistics significantly.

---

galaxy, adopted by MR. In the presence of a finite core radius, a larger asymptotic velocity dispersion is required to produce the same total cross-section as a SIS model.

# 4. Lensing by Early Type Galaxies

## 4.1. Model Calculations

The models presented in the previous section showed that lensing statistics depend sensitively on the evolution of the surface mass density under merging. We will now test a more narrowly specified class of lens models, utilizing additional observational constraints. Specifically, we will employ the composite star/dark halo models described in detail in MR. These models assume that at small radii the stellar surface brightness profile is proportional to the total mass. As observed directly for spirals, the halos then match "smoothly" onto the luminous part of the galaxies, to yield an approximately flat "rotation curve". As discussed in 2.2, this property is expected to be preserved under merging. We normalize the asymptotic velocity dispersions using the inferred parameters of halos at the current epoch and scale them with galaxy luminosity via the Faber-Jackson relation (i.e. $\sigma \propto L^{1/4}$). We adopt a present value of $\sigma_* = 300 \mathrm{km\ s^{-1}}$ for an $L_*$ elliptical and characterize its change under merging by the index $\nu$ from Eq. 4.

Figure 2 is as Figure 1, but for these empirically constrained lens models. For $Q = 2$ the effects of merging are smaller than for SIS models, because a sizeable fraction of the mass inside the critical radii for lensing is provided by the luminous matter, whose structure change is determined by the fundamental plane. The total cross section and the typical image splitting change only little, if merging "moves" galaxies within this plane. Hence, a typical change by a factor of two in the number and mass of galaxies causes a substantially smaller change in the number and mean separation of expected multiple images. For these models $n(\theta)$ always increases st small separations.

The results for $Q = 4$ are qualitatively similar, but more pronounced: models with $\nu = 2/3$ predict even fewer lenses than $Q = 2$, while models with $\nu = 0$ predict tree times as many small separation lenses ($\theta < 1''$) as the no-merging case.

## 4.2.  Comparison with the HST Snapshot Survey

No-evolution models were found to provide an acceptable fit to the incidences of multiple imaging found in the HST Snapshot Survey. Therefore we proceed to ask whether the snapshot data are comparably well fit by the predictions of the merging scenarios. We address this question by calculating the likelihood of the survey observations if a given lens evolution model were correct. This likelihood can be defined as

$$L = \left[ \sum_{i=1}^{M} \ln n(\theta_i) \right] - N \ , \tag{9}$$

where $n(\theta_i)$ is the value of $n(\theta)$ at the separation $\theta_i$ of observed lens systems and $N = \int_0^\infty n(\theta) d\theta$. The sum is taken over the $M = 4$ lenses in the Snapshot Survey (see Figure 2, bottom panel). Following MR, Monte-Carlo simulations then give the probability of obtaining L for a given model.

Table 1 lists the probabilities for various $(Q, \nu)$ pairs. For $Q = 2$ all scenarios with $\nu < 2/3$ produce a probability $\gtrsim 20\%$, and are hence statistically consistent with the HST Snapshot survey data. Only $\nu = 2/3$, which corresponds to Toomre's (1977) "ellipticals form by merger of spirals" scenario is excluded with 95% confidence, primarily because it predicts an $n(\theta)$ that is too small for $\theta > 2''$. For $Q = 4$, only the scenario with $\nu = 1/4$ is still consistent, while $\nu = 0$ and $\nu = 2/3$ are rejected, but for differing reasons: $\nu = 0$ because the models overpredict the number of subarcsecond lenses and $\nu = 2/3$ again because these models underpredict lenses with $\theta > 2''$.

These composite models predict considerably more lenses for $\nu = 2/3$ than simple SIS models (compare Figures 1 and 2). Nevertheless, the case of $\nu = 2/3$, designed to describe the merging of spirals into ellipticals is in conflict with the observed lensing statistics even for a modest amount of merging (i.e. $Q = 2$). Lensing statistics therefore argue against the hypothesis that all or most massive elliptical galaxies in the nearby universe were formed from merging spirals since $z \lesssim 1$. On the other hand, the Snapshot Survey data are consistent with even considerable merging ($Q = 4$), if $\nu = 1/4$, e.g. if the progenitors of present day ellipticals were themselves elliptical galaxies.

## 5. Can Merging "Save" the Cosmological Constant?

It has been argued (Fukugita, Futamase & Kasai 1990; Turner 1990; FT; Bahcall *et al.* 1992; MR; Kochanek 1992) that the statistics of observed gravitational lenses preclude a universe that is dominated by a cosmological constant, $\lambda \gg \Omega_0$. For the no-merging case a large value for $\lambda$ is ruled out because it overpredicts the number of lensed quasars, particularly at small separations ($< 1''$).

These upper limits on $\lambda$ might change, once the no evolution assumption is discarded. However, in light of the discussion in the previous section it is clear that the limits will only tighten: for a given cosmology, merging leads to more small-separation lenses compared to the no-evolution case; such lenses are already over predicted in $\lambda$ dominated cosmologies. Figure 3 shows the predicted lensing distributions $n(\theta)$ for $\lambda = 0.7$ ($\Omega = 0.3$) and $Q = 4$; their associated probabilities are listed in Table 1. In the presence of significant merging, all models with $\lambda = 0.7$ are ruled out with 97.5% confidence.

## 6. Conclusions

We have presented statistical lensing calculations exploring the evolution of both the number and the surface mass density of lens galaxies with cosmological epoch. Employing a simple parameterization ($Q$ and $\nu$) for these two aspects of the evolution, we find that the evolution of surface mass density is at least as important as the number evolution. In almost all scenarios does merging skew $n(\theta)$ towards smaller separations. Using singular isothermal lens models, we showed that without further constraints on the evolution of the surface mass density, the total number of lenses can dramatically increase or decrease by merging.

The existence of a fundamental parameter plane for elliptical galaxies and the fact that stars provide much of the mass in the inner parts of the galaxies, can be used to constrain observationally the evolution of surface mass densities under merging. These external constraints narrow considerably the range of plausible predictions for lensing statistics under merging, making them more similar to the no-evolution prediction.

Only merging scenarios in which galaxies become much more efficient lenses through merging, (such as the "spirals-merge-into-ellipticals" hypothesis), are ruled out for modest amounts of merging ($Q = 2$) on the basis of the HST Snapshot survey data. The merging scenario proposed by Toomre (1977) corresponds approximately to $Q = 2$ and $\nu = 2/3$; this model is rejected at the 95% confidence level by the HST observations.

For larger amounts of merging (Q=4), scenarios in which the mean surface mass density of the merger product is much smaller that its precursor's (leading to inefficient present day lenses) are also ruled out: the abundance of more efficient lenses at earlier epochs would lead to an over prediction of subarcsecond lenses. However, for $\nu = 1/4$, even a large amount of merging, such as suggested by Broadhurst *et al.* (1992), is still consistent with the observed separation distribution of lenses. Note that for $\nu = 1/4$ the differential optical depth at each redshift hardly changes, compared to the null hypothesis. Therefore, even the lens redshift distribution – if it were available – remains unchanged and can not be used as a discriminant.

Finally, we find that the limits on the cosmological constant $\lambda$ derived by assuming no evolution are conservative. $\lambda$-dominated cosmologies are ruled out because they predict a large number of sub-arcsecond lenses, which are not observed. The inclusion of merging only aggravates this discrepancy.

The recent study of Mao and Kochanek (1993) uses similar general techniques, but considers a wide range of changes in the lens population, and relies solely on SIS model galaxies. Our study differs in that it focuses on which evolutionary scenarios are physically most plausible. When considering similar scenarios, their conclusions qualitatively agree with those described above. In particular, they also find that extreme "spiral-to-elliptical" merger scenarios are probably excluded and that merger scenarios do not plausibly allow $\lambda$ dominated models.

While the presently available sample is too small to place strong constraints on a wide variety of merging scenarios, it already excludes a few extreme, but interesting cases of galaxy merging. Should a larger gravitational-lens sample be available in the near future, e.g. from the Digital Sky Survey, a stronger test of merging scenarios could be carried out.

## Acknowledgements

We are grateful to Penny Sackett for a thorough reading of the the manuscript and many helpful suggestions. HWR was supported by a Hubble fellowship (HF-1024.01-91A). ELT and MF are supported in part by the US-Japan Cooperative Science Programme by NSF (INT91-16745) and JSPS. ELT is also supported in part by NSF grant AST90-16533 and NASA grant NAGW-2173. MF also wishes to acknowledge generous support from the Fuji Xerox Corporation.

## Table 1: Models and Probabilities

| $\Omega$ | $\lambda$ | $Q^a$ | $\nu^b$ | $N^c$ | MC%$^d$ |
|---|---|---|---|---|---|
| 1 | 0 | 0 | – | 3.3 | 32 |
| 1 | 0 | 2 | 2/3 | 2.9 | 5.9 |
| 1 | 0 | 2 | 1/4 | 4.6 | 24 |
| 1 | 0 | 2 | 0 | 6.1 | 21 |
| 1 | 0 | 4 | 2/3 | 2.2 | 1.6 |
| 1 | 0 | 4 | 1/4 | 5.5 | 16 |
| 1 | 0 | 4 | 0 | 11.6 | 1.7 |
| 0.2 | 0.8 | 4 | 2/3 | 4.9 | 2.5 |
| 0.2 | 0.8 | 4 | 1/4 | 23 | 0 |
| 0.3 | 0.7 | 4 | 2/3 | 4.2 | 2.8 |
| 0.3 | 0.7 | 4 | 1/4 | 17 | 0 |

$^a$ Amount of merging (see Eq. 1)

$^b$ Change of halo velocity dispersion under merging (see Eq. 5)

$^c$ Expected total number of lenses.

$^d$ Absolute likelihood inferred from Monte-Carlo simulations. The column entry gives the fraction (in %) of Monte-Carlo observation drawn from the model that yielded a lower likelihood than the observed data.

## Figure Captions

Figure 1: Galaxy Merging in the SIS Model

Figure 1 shows the separation distribution of multiply imaged QSO expected for the HST Snapshot Survey, using singular isothermal spheres as a lens model. The thick solid line in each panel represents the no-merging case. The other lines represent merging scenarios of various characters, parameterized by the index $\nu$ (see Eq. 4). The top panel is for a modest amount of merging, $Q = 2$; the center panel is for a larger merging rate, $Q = 4$ (see Equation 1).

Figure 2: Merging of Early Type Galaxies

As Figure 1, but for empirically constrained lens models. The bottom panel shows the predictions of the fiducial lens model used by MR, and a histogram indicating the separations of the four lensed quasars from the HST Snapshot Survey used in the analysis.

Figure 3: Can Merging of Lenses Circumvent the Limits on $\lambda$?

Figure 3 shows the expected separation distribution of multiply imaged QSO expected in the HST Snapshot Survey, as in the "strong merging case" in Figure 2, but with $\Omega = 0.3$ and $\lambda = 0.7$ cosmology. The thick solid line represents the no merging case, the dashed and dash-dotted lines represent the extremes cases for how the halo properties might change under merging.

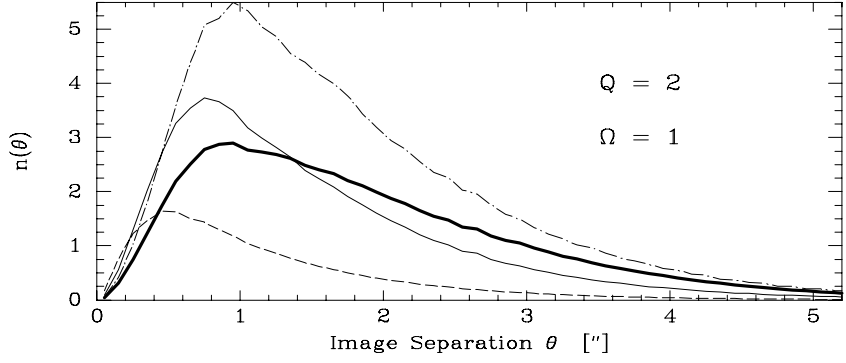

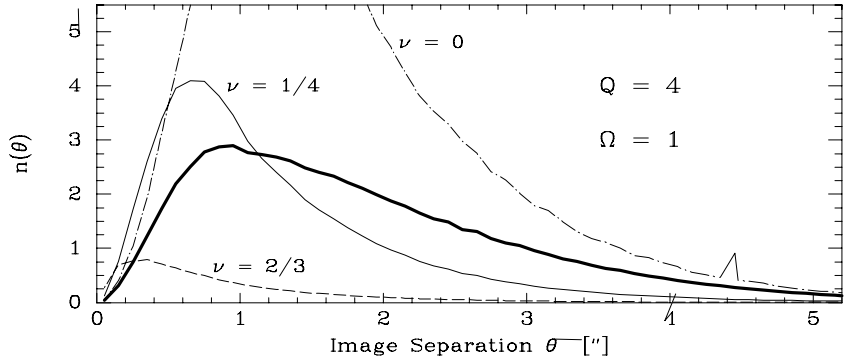

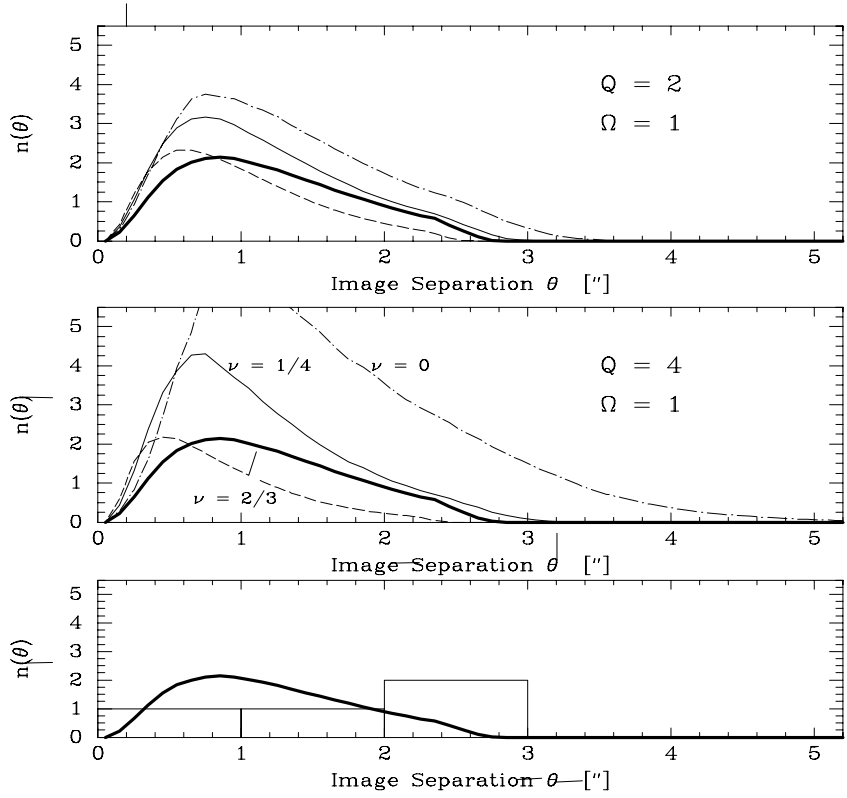

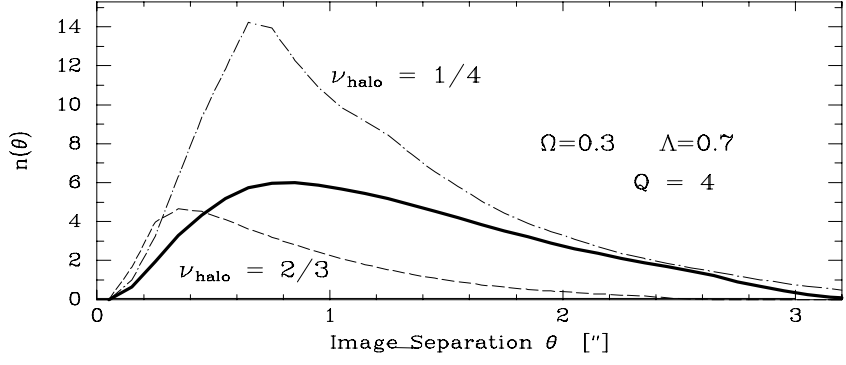